\documentclass[conference]{IEEEtran}
\usepackage{amsmath}
\usepackage{amssymb}
\usepackage{amsthm}
\usepackage{graphicx}
\usepackage{epstopdf}
\usepackage{cite}
\usepackage{algorithm}
\usepackage{algpseudocode}

\IEEEoverridecommandlockouts

\theoremstyle{plain}
\newtheorem{theorem}{Theorem}

\newtheorem{proposition}[theorem]{Proposition}

\theoremstyle{definition}

\theoremstyle{remark}

\DeclareMathOperator{\tr}{tr}
\DeclareMathOperator*{\argmin}{arg\,min}

\begin{document}
\title{Sensor Selection and Power Allocation Strategies for Energy Harvesting Wireless Sensor Networks}
\author{Miguel Calvo-Fullana, Javier Matamoros, and Carles Ant\'on-Haro
\thanks{
This work was partially supported by the Catalan government under grant SGR2014-1567; the Spanish government under grant TEC2013-44591-P (INTENSYV), and grant PCIN-2013-027 (E-CROPS) in the framework of the ERA-NET CHIST-ERA program.

The authors are with the Centre Tecnol\`ogic de Telecomunicacions de Catalunya (CTTC), 08860 Castelldefels, Barcelona, Spain (e-mail: \{miguel.calvo, javier.matamoros, carles.anton\}@cttc.cat).
}
}

\maketitle

\begin{abstract}
In this paper, we investigate the problem of \emph{jointly} selecting a predefined number of energy-harvesting (EH) sensors \emph{and} computing the optimal power allocation. The ultimate goal is to minimize the reconstruction distortion at the fusion center. This optimization problem is, unfortunately, non-convex. To circumvent that, we propose two suboptimal strategies: (i) a \emph{joint} sensor selection and power allocation (JSS-EH) scheme that, we prove, is capable of iteratively finding a stationary solution of the original problem from a sequence of surrogate convex problems; and (ii) a \emph{separate} sensor selection and power
allocation (SS-EH) scheme, on which basis we can identify a sensible sensor selection and analytically find a power allocation policy by solving a convex problem. We also discuss the interplay between the two strategies. Performance in terms of reconstruction distortion, impact of initialization, actual subsets of selected sensors and computed power allocation policies, etc., is assessed by means of computer simulations. To that aim, an EH-agnostic sensor selection strategy, a lower bound on distortion, and an online version of the SS-EH and JSS-EH schemes are derived and used for benchmarking.
\end{abstract}


\IEEEpeerreviewmaketitle

\section{Introduction}
\label{sec:Introduction}

One of the major limiting factors in the lifetime of a Wireless Sensor Network (WSN) is the energy consumption at the sensor nodes. Sensor nodes are typically powered by batteries which can be costly or difficult to replace (e.g., when nodes are deployed in remote locations). To alleviate this problem, \emph{energy harvesting} (EH) has recently emerged as a technology capable of providing self-sustainable operation of those networks. By scavenging energy from solar, thermal, kinetic, electromagnetic or other sources \cite{vullers2010energy}, sensor nodes can extend their operational lifetime. This shifts the reason for the cease of operation from battery depletion to hardware failure.

All this has generated a great deal of research interest in EH techniques and how to effectively exploit such harvested energy (see \cite{ulukus2015energy} and references therein for an overview of current advances). For a \emph{point-to-point} channel, the main focus has been on the derivation of optimal transmission policies in several communication scenarios. For \emph{known} energy and data arrivals (i.e, \emph{offline} optimization) and Gaussian channels, \cite{yang2012optimal} investigates how to minimize the time elapsed until all data packets are transmitted to the destination. These results have been extended to take into account (among others) the impact of finite energy storage \cite{tutuncuoglu2012optimum}, battery leakage \cite{devillers2012general}, communication processing costs \cite{orhan2014energy} and source-channel coding \cite{orhan2014source}. Fading channels have also been considered in \cite{ozel2011transmission,ho2012optimal}, for the derivation of both \emph{online} and \emph{offline} transmission policies. Other communication scenarios with \emph{multiple} EH nodes have been studied in the literature too. This includes the broadcast channel \cite{ozel2012optimal}, the multiple access channel \cite{yang2012optimalMAC}, cooperative transmission schemes \cite{berbakov2014joint}, the interference channel \cite{tutuncuoglu2012sum} and the relay channel \cite{huang2013throughput}.

Besides, current technological advances make it feasible to deploy inexpensive sensors in \emph{large} numbers. In this context, the problem of optimally selecting a subset of sensors to perform a given task naturally arises. This often stems from resource (e.g., bandwidth), interference level or energy consumption constraints, which make massive sensor to Fusion Center (FC) communications barely recommended or simply not possible. While the aforementioned \emph{sensor selection problem} is combinatorial in nature, Joshi and Boyd studied in \cite{joshi2009sensor} a convex relaxation allowing to (approximately) solve the problem with a reasonable computational cost. Other more recent approaches leverage on the inherent sparsity of the problem. For instance, the authors in \cite{jamali2014sparsity} investigate\textemdash both from centralized and distributed standpoints\textemdash strategies aimed to minimize the number of selected sensors subject to a given Mean Square Error (MSE) target. Non-linear measurement models (such as those in source localization and tracking problems) have been considered in \cite{chepuri2015sparsity}, also in a sparsity-promoting framework. Further, the sensor selection problem has also been studied in \cite{liu2015sensor} for correlated measurement noise. From an energy efficiency point of view, the authors in \cite{liu2014energy} used a sparsity-promoting penalty function to discourage the repeated selection of any sensor node in particular (e.g., the most informative ones). By doing so, uneven battery drainage can be prevented. Likewise, the same authors propose in \cite{liu2014optimal} a periodic sensor scheduling strategy which limits the number of times that a sensor can be selected and transmit in a given period of time. In a previous work \cite{calvo2015sensor}, we considered the sensor selection problem in energy harvesting networks. Specifically, we introduced the problem formulation considered in this work, and we derived a separate sensor selection and power allocation scheme. Also, in a related work \cite{calvo2016sparsity}, instead of considering the activation of a predefined number of sensors at each time slot, we relaxed this constraint and developed a globally sparse sensor selection and power allocation scheme.

\subsection{Contribution}

In this paper, in contrast, we investigate the problem of \emph{jointly} selecting a \emph{predefined} number of energy-harvesting sensors and computing the optimal power allocation. The selection is needed due to the reduced number of sensor-to-FC channels. Our goal is to minimize the distortion in the reconstruction of the underlying source at the FC subject to the causality constraints imposed by the EH process.  This is in stark contrast with the approaches in e.g., \cite{liu2014energy} \cite{liu2014optimal} which were EH-agnostic. Unfortunately, the aforementioned optimization problem is not convex. For this reason, we propose \emph{two} suboptimal \emph{offline} strategies. First, the \emph{joint} sensor selection and power allocation (JSS-EH) scheme is capable of finding a stationary solution to the problem (we rigorously prove this) on the basis of a Majorization-Minimization (MM) procedure \cite{lange2000optimization}. The MM procedure allows us to identify a sequence of surrogate (and approximate) convex optimization problems that we iteratively solve. As an alternative, we propose a method to \emph{separately} identify a sensible (and EH-aware) sensor selection and the corresponding power allocation policy. By doing so, the power allocation problem for a \emph{given} sensor selection becomes convex. Hereinafter, this is referred to as the \emph{separate} sensor selection and power allocation (SS-EH) scheme. Very interestingly, the corresponding power allocation policy can be analytically derived and, as we discuss, it can be interpreted as a two-dimensional \cite{orhan2014source} waterfilling solution. Besides, the SS-EH solution turns out to be a suitable initialization to compute in a relatively low number of iterations a \emph{refined} (i.e., with lower distortion) stationary solution to the JSS-EH problem. The contributions in this paper go substantially beyond our initial work in \cite{calvo2015sensor}. Specifically, we propose a new scheme (JSS-EH); for the SS-EH problem, we include a convergence proof and, also, derive an \emph{online} version of both schemes. And, finally, we also discuss the interplay between and conduct an extensive performance assessment of the JSS-EH and SS-EH offline and online schemes by means of computer simulations.

The remainder of this paper is organized as follows. In Section \ref{sec:SystemModel}, we present the signal and system model. In Section \ref{sec:systemModelSS-EH}, we formulate the sensor selection and power allocation problem in an energy-harvesting framework. Sections \ref{sec:JSS-EH} and \ref{sec:SS-EH} are devoted to present the two proposed strategies to compute joint (JSS-EH) and separate (SS-EH) suboptimal solutions to the aforementioned optimization problem, respectively. Next, in Section \ref{sec:NumRes}, we extensively assess the performance of the proposed strategies. Finally, we close the paper by providing some concluding remarks in Section \ref{sec:Conclusions}.

\textit{Notation:} We denote column vectors with bold face letters (e.g., $\mathbf{x}$). When dealing with sensor information, we assume a discrete-time model widely adopted by the information theory community (see for instance \cite{ishwar2005rate,bajwa2007joint,draper2004side}). Given a measurement $y^{(k)}_i[t]$, the superscript denotes the sample index, the subscript denotes the sensor index and the square bracket denotes the time slot index. When dealing with iterations of an algorithm variable, the superscript $(\cdot)^{(k)}$ denotes the iteration number. Moreover, $(\cdot)^T$ denotes the transpose operator and $[\cdot]^{+} \triangleq \max \{\cdot,0\}$.

\section{System Model}
\label{sec:SystemModel}

\begin{figure}[t]
    \centering
    \includegraphics[scale=1]{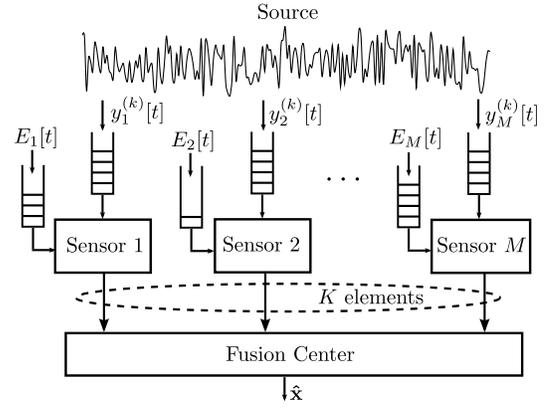} 
    \caption{System model.} 
    \label{fig:systemModel}
\end{figure}

Consider the system model illustrated in Figure \ref{fig:systemModel}, comprising a wireless sensor network composed of $M$ energy harvesting sensor nodes (with index set $\mathcal{M} \triangleq \{1,\ldots,M\}$) and one Fusion Center (FC) deployed to estimate an underlying source $\mathbf{x}\in \mathbb{R}^m$, with $\mathbf{x} \sim \mathcal{N}(0,\mathbf{\Sigma}_x)$. We consider a time-slotted system with $T$ time slots indexed by the set $\mathcal{T} \triangleq \{1,\ldots,T\}$ of duration $T_s$. In time slot $t$, the stationary source $\mathbf{x}$ generates an independent and identically distributed (i.i.d.) large sequence of $n$ samples $\{\mathbf{x}^{(k)}[t]\}_{k=1}^{n}=\left\{\mathbf{x}^{(1)}[t],\ldots,\mathbf{x}^{(n)}[t]\right\}$. As in \cite{joshi2009sensor}, source samples and sensor measurements are related through the following linear model:
\begin{align}
y^{(k)}_i[t]&=\mathbf{a}_i^T\mathbf{x}^{(k)}[t]+w^{(k)}_i[t],   &  \begin{array}{l}
k=1,\dots,n \\
i\in \mathcal{Z}_t,
\end{array}
\end{align}
where $\{w^{(k)}_i[t]\}_{k=1}^{n}$ stands for i.i.d., zero-mean Gaussian observation noise of variance $\sigma_w^2$; vector $\mathbf{a}_i$ gathers the \emph{known} coefficients of the linear model at the $i$-th sensor; and $\mathcal{Z}_t\subseteq \mathcal{M}$ denotes the subset of active (selected) sensors in time slot $t$, with cardinality $|\mathcal{Z}_t|$. The ultimate goal is to reconstruct at the FC the sequence $\{\mathbf{x}^{(k)}[t]\}_{k=1}^{n}$ in each time slot. To that aim, a total of $K \leq M$ \emph{orthogonal} channels are available for sensor-to-FC channel communications. Therefore, the number of sensors selected in each time slot must satisfy $|\mathcal{Z}_t|\leq K$.

In the sequel, we assume separability of source and channel coding. As far as \emph{source} coding is concerned, we adopt a rate-distortion optimal encoder. Assuming a quadratic distortion measure at the FC, the encoded measurements at the sensor nodes can be modeled as a sequence of auxiliary random variables $\{u^{(k)}_i[t]\}_{k=1}^{n}$ \cite{ishwar2005rate}:
\begin{align}
u^{(k)}_i[t]&=\mathbf{a}_i^T\mathbf{x}^{(k)}[t]+w^{(k)}_i[t]+q^{(k)}_i[t],  &  \begin{array}{l}
k=1,\dots,n \\
i\in \mathcal{Z}_t,
\end{array}
\end{align}
with $q^{(k)}_i[t] \sim \mathcal{N}\left(0,\sigma^2_{q_i}[t]\right)$ modeling the i.i.d. encoding noise. The average encoding rate per sample $R_i[t]$ must satisfy the rate-distortion theorem\cite{cover2012elements}, that is,
\begin{align}
R_{i}[t] \geq I(y_i[t];u_i[t])  & = h(u_i[t])-h(u_i[t]|y_i[t]), \nonumber\\
                & = \frac{1}{2}\log\left(1+\frac{\mathbf{a}_i^T \mathbf{\Sigma}_{x}\mathbf{a}_i+\sigma^2_w}{\sigma^2_{q_i}[t]}\right) \label{eq:sourcecodingrate}
\end{align}
for all $i\in \mathcal{Z}_t$. Further, we assume that each \emph{active} sensor encodes its observations at the maximum \emph{channel} rate which is given by the Shannon capacity formula\footnote{For simplicity, we let the number of channel uses per sensor be equal to the number of samples in a time slot.}. Hence we have $R_i[t]=\frac{1}{2}\log(1+h_i[t] p_i[t])$, where $p_i[t]$ and $h_i[t]$ stand for the average transmit power and channel gain, respectively. From this and \eqref{eq:sourcecodingrate}, the variance of the encoding noise reads
\begin{align}
\sigma^2_{q_i}[t]=\frac{\mathbf{a}_i^T \mathbf{\Sigma}_{x}\mathbf{a}_i+\sigma^2_w}{h_i[t] p_i[t]}, \quad i\in\mathcal{Z}_t.
\label{eq:sig2q}
\end{align}
Finally, by means of a Minimum Mean Square Error (MMSE) estimator \cite{kay1993fundamentals} the FC\footnote{The FC collects all measurements and computes the MMSE estimate of the underlying source. Given a general linear model of the form $\mathbf{y}=\mathbf{A}\mathbf{x}+\mathbf{w}$, with $\mathbf{x} \sim \mathcal{N}\left(0,\mathbf{C}_{\mathbf{x}}\right)$, and $\mathbf{w} \sim \mathcal{N}\left(0,\mathbf{C}_{\mathbf{w}}\right)$, the MMSE estimate turns out to be $\mathbf{\hat{x}}=\mathbf{C}_{\mathbf{xy}}\mathbf{C}_{\mathbf{y}}^{-1}\mathbf{y}$ with distortion given by $D_{\text{MMSE}}=\tr(\mathbf{C}_{\mathbf{xy}}\mathbf{C}^{-1}_{\mathbf{y}}\mathbf{C}^T_{\mathbf{xy}}+\mathbf{C_x})^{-1}$, where $\mathbf{C_{y}}=\mathbb{E}[\mathbf{y}\mathbf{y}^T]$ and $\mathbf{C_{xy}}=\mathbb{E}[\mathbf{x}\mathbf{y}^T]$.} reconstructs $\{\mathbf{x}^{(k)}[t]\}_{k=1}^{n}$ from the received codewords $\{u^{(k)}_i[t]\}_{k=1}^{n}$ $i\in\mathcal{Z}_t$. The average (MSE) distortion in time slot $t \in \mathcal{T}$ is given by \cite{kay1993fundamentals}
\begin{align}
D[t]=&
\tr\left(\sum\limits^{M}_{i=1}\frac{z_i[t]}{\sigma^2_w+\sigma^2_{q_i}[t]}\mathbf{a}_i\mathbf{a}_i^T + \mathbf{\Sigma}_x^{-1}\right)^{-1},
\label{eq:Dist}
\end{align}
where $\tr(\cdot)$ denotes the trace operator\footnote{Throughout this paper we adopt the widely accepted notational convention by which the inverse operator precedes the trace operator. That is, $\tr (\mathbf{X})^{-1}$ is understood as $\tr((\mathbf{X})^{-1})$.}, and $\mathbf{z}[t]=[z_1[t],\ldots,z_M[t]]^T$ stands for the sensor selection vector, with $z_i[t]=1$ if  $i\in\mathcal{Z}_t$ and $z_i[t]=0$ otherwise.  By substituting expression \eqref{eq:sig2q} in \eqref{eq:Dist} and defining $\xi_i[t] \triangleq \left(\frac{\mathbf{a}_i^T \mathbf{\Sigma}_{x}\mathbf{a}_i /\sigma^2_w +1}{h_i[t]}\right)$, the distortion can be rewritten as
\begin{align}
D[t]=&
\tr\left(\frac{1}{\sigma^2_w}\sum\limits^{M}_{i=1}\frac{p_i[t]z_i[t]}{p_i[t]+\xi_{i}[t]}\mathbf{a}_i\mathbf{a}_i^T + \mathbf{\Sigma}_x^{-1}\right)^{-1}.
\label{eq:DistXi}
\end{align}

\section{Problem Statement: Sensor Selection and Power Allocation in an Energy Harvesting Framework}
\label{sec:systemModelSS-EH}

Since sensor nodes are capable of harvesting energy from the environment, the average transmit power, $p_i[t]$ in \eqref{eq:DistXi}, is necessarily constrained by the amount of scavenged energy. Hence, in time slot $t \in \mathcal{T}$ we have
\begin{align}
T_{s}\sum_{l=1}^{t}p_{i}[l]\leq &\sum_{l=1}^{t}E_{i}[l], \quad t\in \mathcal{T}, i\in \mathcal{M}.
\label{eq:ECC}
\end{align}
where $E_i[t]$ denotes the energy harvested by the $i$-th sensor node in time slot $t$. In this context, our goal is to \emph{jointly} determine the optimal sensor selection and power allocation strategy that (i) satisfies the above constraints imposed by the energy harvesting process; (ii) selects $K$ sensors in each time slot; and, by doing so, (iii) minimizes the sum distortion \eqref{eq:DistXi} over the $T$ time slots. Accordingly, the optimization problem reads
\begin{subequations}
\begin{align}
   \underset{\mathbf{z}[t],\mathbf{p}[t]}{\text{minimize}}  \quad   &\sum\limits_{t=1}^{T}\tr\left(\frac{1}{\sigma^2_w}\sum\limits^{M}_{i=1}
                        \frac{p_i[t]z_i[t]}{p_i[t]+\xi_{i}[t]}\mathbf{a}_i\mathbf{a}_i^T + \mathbf{\Sigma}_x^{-1}\right)^{-1}
                        \label{eq:OptNonConvexObj}\\
   \text{subject to}
                            \quad & T_{s}\sum_{l=1}^{t}p_{i}[l]\leq \sum_{l=1}^{t}E_{i}[l], \forall t \in \mathcal{T}, \forall i \in \mathcal{M}
                            \label{eq:OptNonConvexECC}\\
                            \quad   & \mathbf{1}^T\mathbf{z}[t]=K, \quad \forall t \in \mathcal{T}  \label{eq:OptNonConvexCount}\\
                            \quad & \mathbf{z}[t] \in \{0,1\}^M, \quad \forall t \in \mathcal{T} \\
                            \quad & \mathbf{p}[t] \geq \mathbf{0}, \quad \forall t \in
                            \mathcal{T} \label{eq:power_ineq}
\end{align}
\label{eq:OptNonConvexEH}
\end{subequations}
where $\mathbf{p}[t]=[p_1[t],\ldots,p_M[t]]^T$ stands for the power allocation vector in a given time slot; $\mathbf{1}$ and $\mathbf{0}$ denote the all-ones and all-zeros vectors of appropriate dimension, respectively; and vector inequality \eqref{eq:power_ineq} is defined elementwise. By introducing the auxiliary vector $\mathbf{s}[t]=[s_1[t],\ldots,s_M[t]]^T$, the optimization problem can be conveniently rewritten as:
\begin{subequations}
\begin{align}
   \underset{\mathbf{z}[t],\mathbf{s}[t],\mathbf{p}[t]}{\text{minimize}}    \quad   &   \sum\limits_{t=1}^{T}\tr\left(
                        \sum\limits_{i=1}^{M}\frac{s_i[t]}{\sigma^2_w}\mathbf{a}_i\mathbf{a}_i^T + \mathbf{\Sigma}_x^{-1}
                        \right)^{-1}        \label{eq:OptRelaxedObjNonConvex}\\
   \text{subject to}
                            \quad & s_i[t] \leq \frac{p_i[t]z_i[t]}{p_i[t]+\xi_{i}[t]}, \forall t \in \mathcal{T}, \forall i \in \mathcal{M} \label{eq:OptRelaxedSValueNonConvex}\\
                            \quad & T_{s}\sum_{l=1}^{t}p_{i}[l]\leq\sum_{l=1}^{t}E_{i}[l], \forall t \in \mathcal{T}, \forall i \in \mathcal{M} \\
                            \quad & \mathbf{1}^T\mathbf{z}[t]=K, \quad \forall t \in \mathcal{T}\\
                            \quad & \mathbf{z}[t] \in \{0,1\}^M, \quad \forall t \in \mathcal{T} \label{eq:OptRelaxedNonConvexBoolean}\\
                            \quad & \mathbf{p}[t] \geq \mathbf{0}, \quad \forall t \in \mathcal{T}  \\
                            \quad & \mathbf{s}[t] \geq \mathbf{0}, \quad \forall t \in \mathcal{T}. \label{eq:OptRelaxedSPositiveNonConvex}
\end{align}
\label{eq:OptRelaxedNonConvex}
\end{subequations}
Clearly, the optimization problems \eqref{eq:OptNonConvexEH} and \eqref{eq:OptRelaxedNonConvex} are equivalent. To see that, note that the objective function is strictly decreasing in $s_i[t]$. Therefore, the optimal solution to problem \eqref{eq:OptRelaxedNonConvex}, namely $\{(z^\star_i[t],s^\star_i[t],p^\star_i[t])\}_{i \in \mathcal{M}, t\in\mathcal{T}}$, must satisfy constraint \eqref{eq:OptRelaxedSValueNonConvex} with equality (since, otherwise, there would be some $s_i[t]>s_i^\star[t]$ for which distortion would be lower, which is a contradiction). That is, we necessarily have $s_i[t] = $ $s^\star_i[t] = p^\star_i[t]z^\star_i[t] / (p^\star_i[t]+\xi_{i}[t]), \forall t \in \mathcal{T}, \forall i \in \mathcal{M}$ which renders the two optimization problems equivalent.

Unfortunately, problem \eqref{eq:OptRelaxedNonConvex} is non-convex due to the Boolean variable $\mathbf{z}[t] \in\{0,1\}^M$ and the product of variables $p_i[t]$ and $z_i[t]$ in constraint \eqref{eq:OptRelaxedSValueNonConvex}. The use of Boolean variables in constraint \eqref{eq:OptRelaxedNonConvexBoolean}, renders the sensor selection problem combinatorial in nature and, in general, NP-hard. To circumvent that, we relax the boolean constraint by letting variable $z_i[t]$ take values in the real-valued interval $[0,1]$ \cite{joshi2009sensor}. The optimization problem now reads
\begin{subequations}
\begin{align}
   \underset{\mathbf{z}[t],\mathbf{s}[t],\mathbf{p}[t]}{\text{minimize}}    \quad   &   \sum\limits_{t=1}^{T}\tr\left(
                        \sum\limits_{i=1}^{M}\frac{s_i[t]}{\sigma^2_w}\mathbf{a}_i\mathbf{a}_i^T + \mathbf{\Sigma}_x^{-1}
                        \right)^{-1}        \\
   \text{subject to}
                            \quad & s_i[t] \leq \frac{p_i[t]z_i[t]}{p_i[t]+\xi_{i}[t]}, \forall t \in \mathcal{T}, \forall i \in \mathcal{M} \label{eq:OptRelaxedSValueNonConvexContinuousZ}\\
                            \quad & T_{s}\sum_{l=1}^{t}p_{i}[l]\leq\sum_{l=1}^{t}E_{i}[l], \forall t \in \mathcal{T}, \forall i \in \mathcal{M} \\
                            \quad & \mathbf{1}^T\mathbf{z}[t]=K, \quad \forall t \in \mathcal{T}\\
                            \quad & \mathbf{z}[t] \in [0,1]^M, \quad \forall t \in \mathcal{T}\\
                            \quad & \mathbf{p}[t] \geq \mathbf{0}, \quad \forall t \in \mathcal{T}  \\
                            \quad & \mathbf{s}[t] \geq \mathbf{0}, \quad \forall t \in \mathcal{T}.
\end{align}
\label{eq:OptRelaxedNonConvexContinuousZ}
\end{subequations}
Still, constraint \eqref{eq:OptRelaxedSValueNonConvexContinuousZ} prevents the optimization problem from being convex. Consequently, one cannot find a global minimizer without resorting to an exhaustive search of the optimization space. Global optimization techniques such as the so-called branch and bound \cite{narendra1977branch} can yield an $\epsilon$-optimal solution but typically exhibit low converge rates and poor scalability with problem dimension.

To alleviate this, we propose two \emph{suboptimal} strategies: an iterative scheme capable of finding a stationary solution to the problem \eqref{eq:OptRelaxedNonConvexContinuousZ} of \emph{jointly} determining the sensor selection and power allocation policies (Section \ref{sec:JSS-EH}); and a method to \emph{separately} identify a sensible sensor selection and a power allocation policy (Section \ref{sec:SS-EH}). Also, we highlight the interplay between these two strategies.
\section{Joint Sensor Selection and Power Allocation with Energy Harvesting (JSS-EH)}
\label{sec:JSS-EH}
Here, we focus on finding a stationary (i.e., at least locally optimal) solution to the problem. To that aim, we resort to a Majorization-Minimization procedure (MM) which is explained below. This technique allows us to iteratively identify a sequence of surrogate (and approximate) convex optimization problems that we attempt to solve.

We start by rearranging the terms of the non-convex constraint \eqref{eq:OptRelaxedSValueNonConvexContinuousZ} as follows:
\begin{align}
s_i[t]p_i[t] -p_i[t]z_i[t] + s_i[t]\xi_{i}[t] \leq 0.
\label{eq:bilinearConstraint}
\end{align}
The terms $f(s_i[t],p_i[t])\triangleq s_i[t]p_i[t]$ and
$g(p_i[t],z_i[t])\triangleq - p_i[t]z_i[t]$, which are bilinear in
the optimization variables, can be alternatively expressed as a
difference of convex functions:
\begin{align}
f(s_i[t],p_i[t])=\frac{1}{2}\left(s_i[t]+p_i[t]\right)^2-\frac{1}{2}\left(s_i[t]^2+p_i[t]^2\right),
\label{eq:spConstraint}
\end{align}
\begin{align}
g(p_i[t],z_i[t])=\frac{1}{2}\left(z_i[t]^2+p_i[t]^2\right)-\frac{1}{2}\left(z_i[t]+p_i[t]\right)^2.
\label{eq:zpConstraint}
\end{align}
In the $k$-th iteration, we obtain a \emph{majorizer} of expression \eqref{eq:bilinearConstraint} by linearizing the concave (second) terms of \eqref{eq:spConstraint} and \eqref{eq:zpConstraint} in the neighborhood of the solution found in the previous iteration ($z_i^{(k)}[t],s_i^{(k)}[t],p_i^{(k)}[t]$), namely
\begin{align}
\bar{f}^{(k)}(s_i[t],p_i[t])&\triangleq\frac{1}{2}\left(s_i[t]+p_i[t]\right)^2-\frac{1}{2}\left(s^{(k)}_i[t]^2+p^{(k)}_i[t]^2\right) \nonumber\\
              &- s^{(k)}_i[t] \left(s_i[t]-s^{(k)}_i[t]\right) \nonumber\\
                  &- p^{(k)}_i[t] \left(p_i[t]-p^{(k)}_i[t]\right),
\label{eq:spConstraintLin}
\end{align}
\vspace{-0.5cm}
\begin{align}
\bar{g}^{(k)}(z_i[t],p_i[t])&\triangleq\frac{1}{2}\left(z_i[t]^2+p_i[t]^2\right)-\frac{1}{2}\left(z^{(k)}_i[t]+p^{(k)}_i[t]\right)^2 \nonumber\\
               &- \left(z^{(k)}_i[t]+p^{(k)}_i[t]\right) \left(z_i[t]-z^{(k)}_i[t]\right) \nonumber\\
               &- \left(z^{(k)}_i[t]+p^{(k)}_i[t]\right) \left(p_i[t]-p^{(k)}_i[t]\right).
\label{eq:zpConstraintLin}
\end{align}
All this results into the following surrogate convex optimization problem for the $k$-th iteration:
\begin{subequations}
\begin{align}
   \underset{\mathbf{z}[t],\mathbf{s}[t],\mathbf{p}[t]}{\text{minimize}}    \quad   &   \sum\limits_{t=1}^{T}\tr\left(
                        \sum\limits_{i=1}^{M}\frac{s_i[t]}{\sigma^2_w}\mathbf{a}_i\mathbf{a}_i^T + \mathbf{\Sigma}_x^{-1}
                        \right)^{-1}        \\
   \text{subject to}
                            \quad & \bar{f}^{(k)}(s_i[t],p_i[t])+\bar{g}^{(k)}(p_i[t],z_i[t]) \label{eq:OptLinearizedConLin}\nonumber\\
                            \quad & \quad\quad\quad +s_i[t]\xi_i[t] \leq 0, \forall t \in \mathcal{T}, \forall i \in \mathcal{M} \\
                            \quad & T_{s}\sum_{l=1}^{t}p_{i}[l]\leq\sum_{l=1}^{t}E_{i}[l], \forall t \in \mathcal{T}, \forall i \in \mathcal{M} \\
                            \quad & \mathbf{1}^T\mathbf{z}[t]=K, \quad \forall t \in \mathcal{T}\\
                            \quad & \mathbf{z}[t] \in [0,1]^M, \quad \forall t \in \mathcal{T} \\
                            \quad & \mathbf{p}[t] \geq \mathbf{0}, \quad \forall t \in \mathcal{T}  \\
                            \quad & \mathbf{s}[t] \geq \mathbf{0}, \quad \forall t \in \mathcal{T}.
\end{align}
\label{eq:OptLinearized}
\end{subequations}
Finally, a stationary point of the original (non-convex) optimization problem \eqref{eq:OptRelaxedNonConvexContinuousZ} can be iteratively found by using Algorithm
\ref{alg:OptAlgorithmNonConvex}.

\begin{algorithm}[t]
    \caption{Joint sensor selection and power allocation.}
    \label{alg:OptAlgorithmNonConvex}
    \begin{algorithmic}[1]
        \State \textbf{Initialize:} Set $k:=0$ and initialize $(\mathbf{z}^{(0)}[t],\mathbf{s}^{(0)}[t],\mathbf{p}^{(0)}[t])$ to a feasible point.
        \State \textbf{Step 1:} Update $\bar{f}^{(k)}$ and $\bar{g}^{(k)}$ according \eqref{eq:spConstraintLin} and \eqref{eq:zpConstraintLin}, respectively.
        \State \textbf{Step 2:} Compute $(\mathbf{z}^{(k+1)}[t],\mathbf{s}^{(k+1)}[t],\mathbf{p}^{(k+1)}[t])$ by solving the optimization problem \eqref{eq:OptLinearized}.
        \State \textbf{Step 3:} Let $k:=k+1$ and go to Step 1 until convergence.
        \State \textbf{Step 4:} Set $\mathbf{z}^\star[t]$ to 1 for the $K$ largest entries in each time slot and 0 otherwise.       
    \end{algorithmic}
\end{algorithm}

\begin{proposition}
Algorithm \ref{alg:OptAlgorithmNonConvex} converges to a stationary solution (a point satisfying the KKT conditions) of the optimization problem \eqref{eq:OptRelaxedNonConvexContinuousZ}.
\end{proposition}
\begin{IEEEproof}
For the ease of notation, let us first collect the vectors of primal variables $\mathbf{z}=[\mathbf{z}[1]^T,\ldots,\mathbf{z}[T]^T]^T$, $\mathbf{s}=[\mathbf{s}[1]^T,\ldots,\mathbf{s}[T]^T]^T$, $\mathbf{p}=[\mathbf{p}[1]^T,\ldots,\mathbf{p}[T]^T]^T$. Let $(\mathbf{z}^{(0)},\mathbf{s}^{(0)},\mathbf{p}^{(0)})$ be a feasible point of the original optimization problem \eqref{eq:OptRelaxedNonConvexContinuousZ}. Since the linearized constraint \eqref{eq:OptLinearizedConLin} is an upper bound on the original constraint \eqref{eq:OptRelaxedSValueNonConvexContinuousZ}, it follows that the feasible set of the surrogate problem \eqref{eq:OptLinearized} at iteration $k$, is contained in the feasible set of the original problem \eqref{eq:OptRelaxedNonConvexContinuousZ}. Hence, all iterates are feasible.

Now, solving the optimization problem \eqref{eq:OptLinearized} at iteration $k$ leads to a solution $(\mathbf{z}^{(k+1)},\mathbf{s}^{(k+1)},\mathbf{p}^{(k+1)})$ satisfying  $\sum_{t=1}^{T}\tr(\sum_{i=1}^{M}(s^{(k+1)}_i[t]/ \sigma^2_w)\mathbf{a}_i\mathbf{a}_i^T + \mathbf{\Sigma}_x^{-1})^{-1} \leq \sum_{t=1}^{T}\tr(\sum_{i=1}^{M}(s^{(k)}_i[t] / \sigma^2_w)\mathbf{a}_i\mathbf{a}_i^T + \mathbf{\Sigma}_x^{-1})^{-1}$. This follows from the fact that, by definition, the linearization is tight at the point $(\mathbf{z}^{(k)},\mathbf{s}^{(k)},\mathbf{p}^{(k)})$, and that by convexity of problem \eqref{eq:OptLinearized} we have $(\mathbf{z}^{(k+1)},\mathbf{s}^{(k+1)},\mathbf{p}^{(k+1)})=(\mathbf{z}^{(k)},\mathbf{s}^{(k)},\mathbf{p}^{(k)})$ if $(\mathbf{z}^{(k)},\mathbf{s}^{(k)},\mathbf{p}^{(k)})$ is a minimizer of the $k+1$ iteration. Thus, the sequence of objective functions generated by Algorithm \ref{alg:OptAlgorithmNonConvex} is nonincreasing and bounded, therefore it converges. Denote the primal variables at this point by $(\mathbf{z}^{\star},\mathbf{s}^{\star},\mathbf{p}^{\star})$, and the corresponding dual variables by $\boldsymbol{\lambda}^{\star}$. Since problem \eqref{eq:OptLinearized} satisfies Slater's condition, its Lagrangian has a saddle point in $\bigl((\mathbf{z}^{\star},\mathbf{s}^{\star},\mathbf{p}^{\star}),\boldsymbol{\lambda}^{\star}\bigr)$.

However, since the linearization is tight at the point $(\mathbf{z}^{\star},\mathbf{s}^{\star},\mathbf{p}^{\star})$, the gradients in the KKT conditions of problems \eqref{eq:OptRelaxedNonConvexContinuousZ} and \eqref{eq:OptLinearized} match. To see this, let $s_i[t]=s^\star_i[t]=s^{(k)}_i[t]$ and  $p_i[t]=p^\star_i[t]=p^{(k)}_i[t]$  in expressions \eqref{eq:spConstraintLin} and \eqref{eq:zpConstraintLin} (they become equivalent to \eqref{eq:spConstraint} and \eqref{eq:zpConstraint}, respectively). Therefore, Algorithm \ref{alg:OptAlgorithmNonConvex} converges to a KKT point of the optimization problem \eqref{eq:OptRelaxedNonConvexContinuousZ}.
\end{IEEEproof}

\subsection{Remarks}
A few considerations are in line. First, in order to select a subset of sensors after convergence, the (relaxed) solution $\mathbf{z}^\star[t] \in [0,1]^M$ must be forced to take Boolean values again, namely $\mathbf{z}^\star[t] \in \{0,1\}^M$. To that aim, the $\mathbf{z}^\star[t]$ vectors are cropped to their $K$ largest entries. After that, however, we do not recompute the associated power allocation. This, however, has a negligible impact on performance since, as discussed in the numerical results section, typically just $K$ entries in vector $\mathbf{z}^\star[t]$ are numerically close to 1, whereas the rest are approximately 0.

The second consideration is that, being the problem non-convex and Algorithm \ref{alg:OptAlgorithmNonConvex} iterative, the stationary solution at which it converges depends on the initialization (and so does performance). Hence, providing it with a suitable initialization is crucial. This will be further discussed in the next section.

Finally, note that one full convex optimization problem has to be solved in each iteration, the computational burden of which might not be negligible (in particular if the number of constraints is large). Therefore, special attention should be paid to the number of iterations needed and the increased computational burden that this entails (for some initializations, convergence can be particulary slow, see numerical results).
\section{Separate Sensor Selection and Power Allocation with Energy Harvesting (SS-EH)}
\label{sec:SS-EH}

Here, we depart from the \emph{iterative} scheme presented in the previous section. Instead, we propose a lower complexity \emph{one shot} approach. Specifically, we propose to determine the subset of active sensors first, and then compute the optimal power allocation policy for such selection. Interestingly, the latter turns out to be a convex (and, thus, easily solvable) problem.

\subsection{Optimal Power Allocation for a Given Sensor Selection}

\begin{figure}[t!]
    \centering
    \includegraphics[scale=1.25]{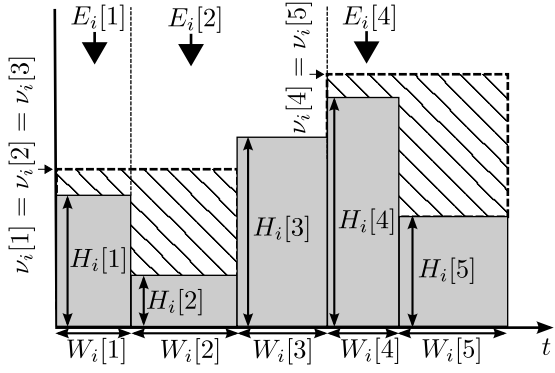}
    \caption{Two-dimensional directional waterfilling for a sensor $i \in \mathcal{M}$ in a scenario with $T=5$ time slots and energy arrivals in time slots 1,2 and 4.}
    \label{fig:WF}
\end{figure}

For a given subset $\{\mathcal{Z}_t\}_{t \in \mathcal{T}}$ of active sensors in \emph{each} time slot, the resulting optimization problem \eqref{eq:OptRelaxedNonConvex} reads
\begin{subequations}
\begin{align}
   \underset{\mathbf{s}[t],\mathbf{p}[t]}{\text{minimize}}  \quad   &   \sum\limits_{t=1}^{T}\tr\left(
                        \sum\limits_{i\in\mathcal{Z}_t}\frac{s_i[t]}{\sigma^2_w}\mathbf{a}_i\mathbf{a}_i^T + \mathbf{\Sigma}_x^{-1}
                        \right)^{-1}        \label{eq:OptRelaxedObj}\\
 \text{subject to}
                            \quad & s_i[t] \leq \frac{p_i[t]}{p_i[t]+\xi_{i}[t]}, \forall t \in \mathcal{T}, \forall i \in \mathcal{M} \label{eq:OptRelaxedSValue}\\
                            \quad & T_{s}\sum_{l=1}^{t}p_{i}[l]\leq\sum_{l=1}^{t}E_{i}[l], \forall t \in \mathcal{T}, \forall i \in \mathcal{Z}_t \label{eq:EH_constraint}\\
                            \quad & \mathbf{p}[t] \geq \mathbf{0}, \quad \forall t \in \mathcal{T}  \\
                            \quad & \mathbf{s}[t] \geq \mathbf{0}, \quad \forall t \in \mathcal{T} \label{eq:OptRelaxedSPositive}
\end{align}
\label{eq:OptRelaxed}
\end{subequations}
where, clearly, the sensor selection vector $\mathbf{z}[t]$ has been removed from the problem formulation. Since the objective function \eqref{eq:OptRelaxedObj} is convex and the constraints \eqref{eq:OptRelaxedSValue}-\eqref{eq:OptRelaxedSPositive} define a convex feasible set, the resulting optimization problem \eqref{eq:OptRelaxed} is convex and therefore has a global minimizer \cite{boyd2009convex}. By satisfying the Karush-Kuhn-Tucker (KKT) conditions, we can identify the necessary and sufficient conditions for optimality. Specifically, the Lagrangian of \eqref{eq:OptRelaxed} is given by
\begin{align}
\mathcal{L}&=\sum\limits_{t=1}^{T}\tr\left(\sum\limits_{i \in \mathcal{Z}_t}\frac{s_i[t]}{\sigma^2_w}\mathbf{a}_i\mathbf{a}_i^T +
        \mathbf{\Sigma}_x^{-1}\right)^{-1} \nonumber\\
        &+\sum\limits_{t=1}^{T}\sum\limits_{i \in \mathcal{Z}_t}\lambda_i[t] \left(s_i[t] -
        \frac{p_i[t]}{p_i[t]+\xi_{i}[t]}\right) \nonumber\\
        &+\sum\limits_{t=1}^{T}\sum\limits_{i \in \mathcal{Z}_t}\beta_i[t]
        \left(T_{s}\sum_{l=1}^{t}p_{i}[l]-\sum_{l=1}^{t}E_{i}[l]\right)\nonumber\\
        &-\sum\limits_{t=1}^{T}\sum\limits_{i \in \mathcal{Z}_t}\eta_i[t]p_i[t]
          -\sum\limits_{t=1}^{T}\sum\limits_{i \in \mathcal{Z}_t}\theta_i[t]s_i[t],
\label{eq:OptRelaxedLagrangian}
\end{align}
where $\lambda_i[t] \geq 0$, $\beta_i[t] \geq 0$, $\eta_i[t] \geq 0$ and $\theta_i[t] \geq 0$ are the corresponding dual variables. By taking the derivative of the Lagrangian with respect to $p_i[t]$, we get
\begin{align}
\frac{\partial\mathcal{L}}{\partial p_i[t]} =
\frac{-\lambda_i[t]\left(p_i[t]+\xi_i[t]\right)+\lambda_i[t]p_i[t]}{\left(p_i[t]+\xi_i[t]\right)^2}
+T_s\sum_{l=t}^{T}\beta_i[l]-\eta_i[t]\nonumber
\end{align}
By letting $\frac{\partial\mathcal{L}}{\partial p_i[t]} =0$ and applying the complementary slackness condition $\eta_i[t]p^\star_i[t]=0$, the optimal power allocation $p^{\star}_i[t]$ follows:
\begin{align}
p^{\star}_i[t]=\sqrt{\frac{\xi_i[t]\lambda_i[t]}{T_s}}\left[
\frac{1}{\sqrt{\sum_{l=t}^{T}\beta_i[l]}}-\sqrt{\frac{\xi_i[t] T_s}{\lambda_i[t]}}
\right]^+,
\end{align}
where $[\cdot]^{+}=\max \{\cdot,0\}$. This solution can be interpreted as the two-dimensional  directional waterfilling shown in Figure \ref{fig:WF}. For an arbitrary sensor $i$, each time slot is associated to a rectangle of solid material of width $W_i[t] \triangleq \sqrt{\xi_i[t]\lambda_i[t]/ T_s}$ and height $H_i[t] \triangleq \sqrt{\xi_i[t] T_s/\lambda_i[t]}$. Right-permeable walls are placed at each time slot with an energy arrival ($t=1, 2, 4$), this accounting for the causality of energy consumption. Water is then poured up to a waterlevel given by $\nu_i[t] \triangleq 1\big/\sqrt{\sum_{l=t}^{T}\beta_i[l]}$. Finally, the corresponding power allocation is given by the area of water above the solid rectangle.

Next, the derivative of the Lagrangian w.r.t. $s_i[t]$ yields
\begin{align}
\frac{\partial\mathcal{L}}{\partial s_i[t]} & =-
\tr\left(\left(\sum\limits_{j \in \mathcal{Z}_t}\frac{s_j[t]}{\sigma^2_w}\mathbf{a}_j\mathbf{a}_j^T + \mathbf{\Sigma}_x^{-1}\right)^{-2} \nonumber
\left(\mathbf{a}_i\mathbf{a}_i^T\right)\right)
\\ & +\lambda_i[t]-\theta_i[t].
\label{Lag_s_t}
\end{align}
Unfortunately, from \eqref{Lag_s_t} no closed-form expression can be found for $s_i^{\star}[t]$. Hence, $s_i[t]$ will be iteratively updated by means of the projected gradient method \cite{bertsekas1999nonlinear}.

Algorithm \ref{alg:OptAlgorithm} summarizes the proposed procedure for the computation of the optimal power allocation. Specifically, we use an Uzawa update step \cite{arrow1958studies} to find the optimal primal-dual saddle point of the optimization problem \eqref{eq:OptRelaxed}. In this way, at each iteration we do an exact
minimization of the power allocation $\{p_i[t]\}$ while we iteratively update both the auxiliary $\{s_i[t]\}$ and the dual $\{\lambda_i[t]\}$ variables. Convergence of Algorithm 2 is trivially satisfied by the Arrow-Hurwicz-Uzawa method, as it is shown next.

\begin{proposition}
Algorithm \ref{alg:OptAlgorithm} converges to the global minimum of the optimization problem \eqref{eq:OptRelaxed}.
\end{proposition}
\begin{IEEEproof}
For the ease of notation, let us first collect the vectors of primal variables $\mathbf{s}=[\mathbf{s}[1]^T,\ldots,\mathbf{s}[T]^T]^T$, $\mathbf{p}=[\mathbf{p}[1]^T,\ldots,\mathbf{p}[T]^T]^T$, and let $\boldsymbol{\lambda}$ be the vector of all dual variables. Since problem \eqref{eq:OptRelaxed} satisfies Slater's condition, the Lagrangian \eqref{eq:OptRelaxedLagrangian} of this optimization problem satisfies the saddle-point property, namely
\begin{align}
\min\limits_{\mathbf{s},\mathbf{p}} \max\limits_{\boldsymbol{\lambda}} \mathcal{L}(\mathbf{s},\mathbf{p},\boldsymbol{\lambda})= \max\limits_{\boldsymbol{\lambda}} \min\limits_{\mathbf{s},\mathbf{p}} \mathcal{L}(\mathbf{s},\mathbf{p},\boldsymbol{\lambda}).
\end{align}
Let us define $\mathbf{p}^\star(\boldsymbol{\lambda}) \triangleq \argmin_{\mathbf{p}}\mathcal{L}(\mathbf{s},\mathbf{p},\boldsymbol{\lambda})$, then for Algorithm \ref{alg:OptAlgorithm} to converge by the Uzawa method \cite{arrow1958studies}, the saddle-point property must also be satisfied given $\mathbf{p}^\star(\boldsymbol{\lambda})$ for all $\mathbf{s}$, that is
\begin{align}
\min\limits_{\mathbf{s}} \max\limits_{\boldsymbol{\lambda}} \mathcal{L}(\mathbf{s},\mathbf{p}^\star(\boldsymbol{\lambda}),\boldsymbol{\lambda})= \max\limits_{\boldsymbol{\lambda}} \min\limits_{\mathbf{s}} \mathcal{L}(\mathbf{s},\mathbf{p}^\star(\boldsymbol{\lambda}),\boldsymbol{\lambda}).
\end{align}
This is equivalent to
\begin{align}
\min\limits_{\mathbf{s}} \min\limits_{\mathbf{p}} \max\limits_{\boldsymbol{\lambda}} \mathcal{L}(\mathbf{s},\mathbf{p},\boldsymbol{\lambda})= \min\limits_{\mathbf{s}} \max\limits_{\boldsymbol{\lambda}} \min\limits_{\mathbf{p}} \mathcal{L}(\mathbf{s},\mathbf{p},\boldsymbol{\lambda}),
\end{align}
which is to say that the Lagrangian \eqref{eq:OptRelaxedLagrangian} must have a saddle point in $(\mathbf{p},\boldsymbol{\lambda})$ for all $\mathbf{s}$, namely
\begin{align}
\min\limits_{\mathbf{p}} \max\limits_{\boldsymbol{\lambda}} \mathcal{L}(\mathbf{s},\mathbf{p},\boldsymbol{\lambda})=
\max\limits_{\boldsymbol{\lambda}} \min\limits_{\mathbf{p}} \mathcal{L}(\mathbf{s},\mathbf{p},\boldsymbol{\lambda}).
\label{eq:ProofAlg2SaddlePoint}
\end{align}
This corresponds to solving optimization problem \eqref{eq:OptRelaxed} with a fixed value of $\mathbf{s}$. Since this problem also satisfies Slater's condition, the saddle-point property \eqref{eq:ProofAlg2SaddlePoint} is satisfied. Therefore, convergence of the Algorithm \ref{alg:OptAlgorithm} follows by convergence of the inexact Uzawa algorithm \cite[Theorem 3.1]{bramble1997analysis}.
\end{IEEEproof}

\begin{algorithm}[t]
    \caption{Optimal power allocation for a given sensor selection.}
    \label{alg:OptAlgorithm}
    \begin{algorithmic}[1]
        \State \textbf{Initialize:} $\{\lambda_i[t]\}:=0$, $\{s_i[t]\}:=0$, select $\epsilon$.
        \State \textbf{Step 1:} For all $t \in \mathcal{T}$ and $i \in \mathcal{Z}_t$, update primal variables.
            \State $s^{(k+1)}_i[t]:=\left[s^{(k)}_i[t]-\epsilon\left(\lambda^{(k)}_i[t]- \right. \right.$\par
            \hskip\algorithmicindent
            $
            \left.\tr\left(\left(\sum\limits_{j \in \mathcal{Z}_t}\frac{s_j^{(k)}}{\sigma^2_w}[t]\mathbf{a}_j\mathbf{a}_j^T+\mathbf{\Sigma}_x^{-1}\right)^{-2}
            \left(\mathbf{a}_i\mathbf{a}_i^T\right)\right)\right]^+
            $
            \State $p^{(k+1)}_i[t]  :=
            \sqrt{\frac{\xi_i[t]\lambda^{(k)}_i[t]}{T_s}}\left[\frac{1}{\sqrt{\sum_{l=t}^{T}\beta_i[l]}}-\sqrt{\frac{\xi_i[t] T_s}{\lambda^{(k)}_i[t]}}\right]^+$
        \State \textbf{Step 2:}  For all $t \in \mathcal{T}$ and $i \in \mathcal{Z}_t$, update dual variable.
            \State $\lambda^{(k+1)}_i[t]  :=
            \left[\lambda^{(k)}_i[t]+\epsilon\left(s^{(k+1)}_i[t] - \frac{p^{(k+1)}_i[t]}{p^{(k+1)}_i[t]+\xi_i[t]}\right)\right]^+$
        \State \textbf{Step 3:} Go to Step 1 until termination condition is met.
    \end{algorithmic}
\end{algorithm}

\subsection{EH-aware Sensor Selection}
\label{sec:EW_aware_SS}

For a system \emph{without} energy harvesting sensors, Joshi and Boyd \cite{joshi2009sensor} propose to compute the sensor selection policy by solving the convex program
\begin{align}
   \underset{\mathbf{z}}{\text{minimize}}   \quad   &   \tr\left(\sigma^{-2}_w\sum\limits^{M}_{i=1}z_i
                                    \mathbf{a}_i\mathbf{a}_i^T + \mathbf{\Sigma}_x^{-1}\right)^{-1}         \label{eq:OptOriginal}\\
   \text{subject to}                    \quad   & \mathbf{1}^T\mathbf{z}=K, \enskip \mathbf{z} \in [0,1]^M.      \nonumber
\end{align}
and constructing the selection sets $\{\mathcal{Z}_t\}_{t \in \mathcal{T}}$ from the $K$ largest elements in the \emph{sensor selection} vector $\mathbf{z}^{\star}[t]$ (in our case, the \emph{same} subset of sensors for \emph{all} time slots). This indexed family of sets $\{\mathcal{Z}_t\}_{t \in \mathcal{T}}$ can then be used to solve the optimization problem \eqref{eq:OptRelaxed} in the preceding subsection and, by doing so, compute the (associated) optimal power allocation.

This EH-agnostic policy\footnote{Note that it only takes into account the impact of $\mathbf{a}_i$, i.e., the set of coefficients in the linear observation model of each node.} might select sensors which do not have any harvested energy yet. To circumvent that, we propose a (heuristic) EH-aware sensor selection policy. First, we let $\mathcal{Z}_t=\mathcal{M}$ and solve problem \eqref{eq:OptRelaxed}. Clearly, ${s}_i[t]$ in \eqref{eq:OptRelaxed} plays the same role as $z_i$ does in \eqref{eq:OptOriginal}, namely, it weights the contribution of each sensor to the resulting distortion. Motivated by this, an intuitive selection rule consists
in choosing for \emph{each} time slot $t$ the $K$ largest elements in vector $\mathbf{s}^\star[t]$. With the indexes of these elements, we form the new selection sets $\{\mathcal{Z}_t\}_{t \in \mathcal{T}}$. And by solving problem \eqref{eq:OptRelaxed} again with this new indexed family of sets, we obtain the corresponding
optimal power allocation. The main difference is that, now, $\mathbf{s}^\star[t]$ takes into account not only the impact of $\mathbf{a}_i$ but also the actual energy arrivals via the energy causality constraint \eqref{eq:EH_constraint}.

\subsection{Remarks}
As discussed earlier, the computational complexity of the \emph{separate} sensor selection and power allocation approach (SS-EH), which is one-shot, is lower than that of the \emph{joint} one (JSS-EH) presented in the previous section, which is iterative. However, no guarantee on the optimality of either solution can be
given. Still, if the former is initialized with the solution to the latter, it will be capable of \emph{refining} it. To recall, we proved that JSS-EH always converges to a stationary solution of the original problem. Therefore, the resulting distortion after convergence will necessary be lower (i.e., a \emph{refined} solution). In general, the solution to the separate optimization problem turns out to be a suitable initialization for the JSS-EH scheme.
\subsection{Online SS-EH Strategy}
The proposed SS-EH scheme requires \emph{non-causal} knowledge on energy arrivals. Here, instead, we introduce a more realistic \emph{online} version just requiring \emph{causal} knowledge\footnote{Likewise, an online version can be derived for the JSS-EH scheme. However, we focus on SS-EH, for brevity. Nonetheless, numerical results are provided in Section \ref{sec:NumRes} for both schemes.}.

Inspired by \cite{ozel2011transmission}, a \emph{myopic} online policy can be computed as follows. Assume for a moment that, after harvesting some energy in the initial time slot, no additional energy is harvested by the sensors. Hence, we let $E_i[1]>0$ and $E_i[2]=\cdots=E_i[T]=0$ for all $i$, and solve the sensor selection
and power allocation problem \eqref{eq:OptRelaxed} for $t=1,\ldots,T$.

In the absence of knowledge on future energy arrivals, this is also a sensible approach. After all, the reconstruction distortion is minimized if no additional energy is harvested. Let $t_o\leq T$ denote the next time slot in which some energy is harvested by an arbitrary sensor $i_o$ i.e., $E_{i_o}[t_o]>0$). For the preceding time slots (i.e., $t=1,\ldots,t_o-1$), we impose that the subsets of active sensors and the power allocation just computed remain unchanged. Hence, the remaining (unspent) energy at the beginning of time slot $t_o$ reads $E_i^u[t_o]=\sum_{t=1}^{t_o-1}E_i[t]-T_s\sum_{t=1}^{t_o-1}p_i[t]$ for \emph{all} $i$. Further, we let
\begin{align}
E_{i}[t_o]:=	
\begin{cases}
	E_{i}^u[t_o] + E_{i}[t_o]		&	\text{if } i= i_o, \\
	E_{i}^u[t_o]				&	\text{if } i\neq i_o,
\end{cases}
\end{align}
$E_i[t_o+1],\ldots,E_i[T]:=0$ for all $i$ and, then, we compute the sensor selection and power allocation for $t=t_o,\ldots,T$, that is, for all subsequent time slots. This procedure is iterated until all energy arrivals have been accounted for. The interesting property of such scheme is its ability to adjust (recompute) the remaining subsets of active sensors and power allocations whenever some additional energy is harvested. By doing so, the additional (and causal) knowledge on energy arrivals is effectively exploited. This myopic policy, however, has side effect: it tends to generate \emph{conservative} power allocation patterns. That is, it tends to shift power allocation towards the
\emph{end} of the observation period (i.e., time slot $T$). To recall, when the power allocation is recomputed after harvesting some energy, the working assumption is that no additional energy will be harvested anymore. Consequently, the algorithm tends to spend energy very slowly, to make sure that for each sensor some energy is left for data transmission for the whole observation window (since, it can be shown that transmitting over longer time periods results into lower distortion).

\section{Numerical Results}
\label{sec:NumRes}

In this section, we assess the performance of the two proposed energy harvesting-aware sensor selection and power allocation strategies. Unless otherwise stated, the algorithm to solve the JSS-EH problem is initialized with the solution of the SS-EH problem. As a benchmark, we use the EH-agnostic policy (SS) proposed in \cite{joshi2009sensor} and succinctly described in Section \ref{sec:EW_aware_SS}. For simulation purposes, we consider linear combination coefficients (which, to recall, are held fixed for all time slots) given by $\mathbf{a}_i\sim\mathcal{N}(0,\mathbf{I}/\sqrt{m})$, with $m=5$ for the underlying source. Energy arrivals $E_{i}[t]$ are modeled as Poisson processes of intensity rate $\mu$ and $|E_{i}[t]|=E$. Further, we assume static (i.e., non-fading) sensor-to-FC channels\footnote{This is a reasonable assumption for static wireless sensor networks.}.

\subsection{Subsets of Active Sensors}

\begin{figure}[t]
    \centering
    \includegraphics[width=\columnwidth]{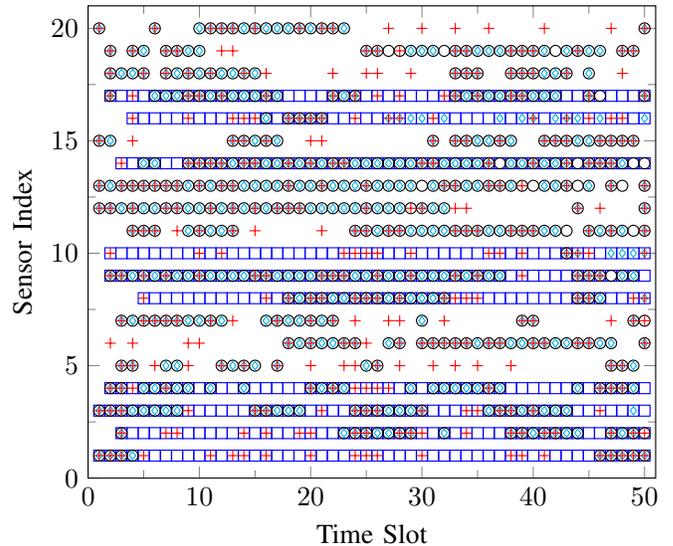}
    \caption{Sensor selection policies. Energy arrivals are denoted by $+$. The sensors selected by the SS, SS-EH and JSS-EH policies are denoted by {\tiny $\square$}, $\circ$ and $\diamond$, respectively ($M=20$, $T=50$, $K=10$, $\mu=0.5$, $\sigma_w^2=0.1$).}
    \label{fig:SelectionPolicies}
\end{figure}

In Figure \ref{fig:SelectionPolicies}, we depict an individual realization of \emph{subsets} of active sensors associated to the JSS-EH, SS-EH and SS strategies. Specifically, a marker is shown whenever a particular sensor belongs to the subset of selected sensors \emph{and} some transmit power is allocated to it (i.e.,
$p_i[t]> 0$)\footnote{Notice that the former does not necessary imply the latter for the SS strategy until some energy is harvested by each sensor.}. The number of selected sensors in each time slot is set to $K=10$ (out of $M=20$).

\begin{figure}[t]
    \centering
    \centering
    \includegraphics[width=\columnwidth]{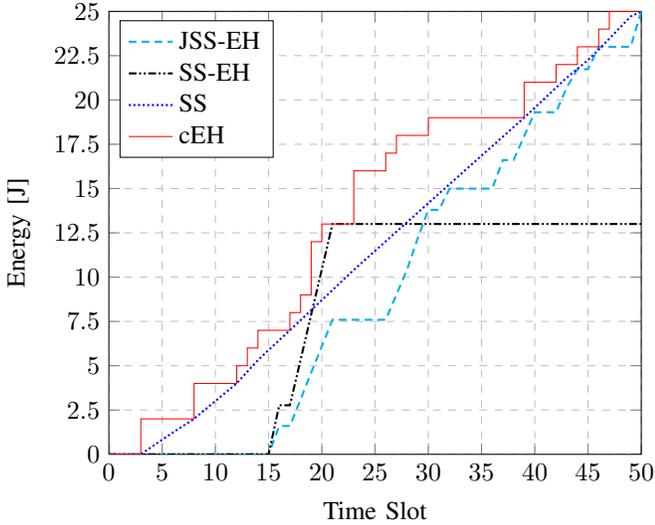}    
    \caption{Power allocation policies corresponding to sensor 16 in Fig. \ref{fig:SelectionPolicies} for the Joint (JSS-EH) and Separate (SS-EH) EH-aware Sensor Selection and Power Allocation strategies, and the EH-agnostic Sensor Selection (SS) one. The cumulative energy harvesting (cEH) curve is shown as a reference.}
    \label{fig:SelectionPoliciesSensor16}
\end{figure}

As discussed earlier, the SS strategy selects the \emph{same} subset of sensors for all time slots, that is, irrespectively of energy arrivals. Specifically, it tends to select the sensors with the most informative observations according to the generated $\mathbf{a}_i$ vectors. On the contrary, the active sensors resulting from the
proposed SS-EH and JSS-EH strategies vary from time slot to time slot (since they \emph{do} take into account energy arrivals). This results into a more efficient use of the available energy.

Interestingly enough, the subsets of active sensors for the SS-EH and JSS-EH strategies are very similar. The most notable difference is sensor 16, which remains inactive after time slot 21 for SS-EH, whereas it is included in the scheduling pattern of JSS-EH until the very last time slot. As discussed earlier, JSS-EH manages to
\emph{refine} the solution of the SS-EH problem and, hence, no radical changes can, in principle, be expected. However, selectively introducing some adjustments may have a considerable impact on the resulting distortion (see Section \ref{sec:dist_performance} ahead). To illustrate this, Figure \ref{fig:SelectionPoliciesSensor16} shows the power allocation for sensor 16 associated to the three strategies. Since the SS-EH strategy does not select this sensor
after time slot 21, part of the harvested energy is wasted (i.e., not used for transmission). This stems from the fact that the actual selection rule (based on the $\mathbf{s}[t]$ values) yet more sophisticated than a EH-agnostic one is, in fact, heuristic. The JSS-EH strategy fixes this inefficiency by properly adjusting the power allocation policy. This allows to schedule sensor 16 after time slot 21 too and, by doing so, consume the energy that is harvested after that time slot.

\begin{figure}[t]
    \centering
    \includegraphics[width=\columnwidth]{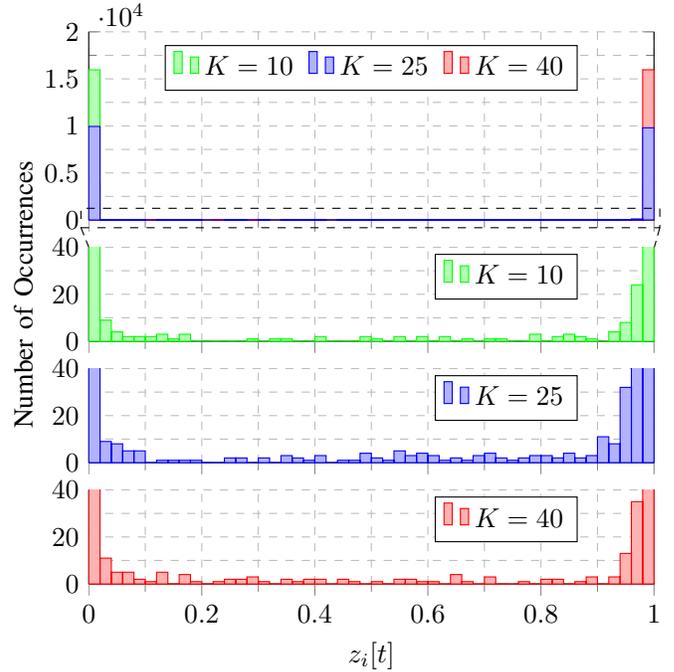}
\caption{Histogram of the selection variable $z_i[t]$ after convergence of the JSS-EH scheme (top) and zoomed-in area with details (bottom). Results are shown for a total of 20 independent runs with random initializations ($M=50$, $T=20$, $\mu=1$, $\sigma_w^2=0.1$). } \label{fig:Histogram}
\end{figure}

\subsection{Impact of Cropping the Selection Vector}
To recall, in order to effectively select a subset of sensors the JSS-EH scheme forces (crops) $\mathbf{z}^\star[t]$ to 1 for the $K$ largest entries in each time slot (and 0 otherwise). However, we argued, there is no need to recompute corresponding power allocation. Figure \ref{fig:Histogram}, which shows a histogram of
the $z_i[t]$ variables after convergence (and right before cropping, for $20$ independent runs with random initializations and repeated for a different number of selected sensors, $K$), evidences why: with high probability, those values already lie in a close neighborhood of 0 or 1. Also, as the figure reveals, for the intermediate values of $z_i[t]$ in the histogram (i.e., those in between 0 and 1) the actual $K$ parameter setting has virtually no impact. Take for example the case $K=25$, since the percentage of active sensors is $K/M=25/50=50\%$ the bar in 1 is of (roughly) the same height as that in 0. The zoomed-in area reveals that only a small percentage of values lie in between 0 and 1: $1.72\%$ (or $342$ out of $20{,}000$) in the interval $(0.01,0.99)$; or $0.38\%$ (or $76$ out of $20{,}000$) in the interval $(0.1,0.9)$. The constraint $\mathbf{1}^T\mathbf{z}[t]=K$ thus implies that, for the largest $K$ values in each time slot (and only those ones), we have $z_i[t]\thickapprox 1$. Therefore, the impact of not recomputing the power allocation for such a reduced subset of $K$ sensors is negligible.

\subsection{Distortion Performance}
\label{sec:dist_performance}

\begin{figure}[t]
    \centering
    \includegraphics[width=\columnwidth]{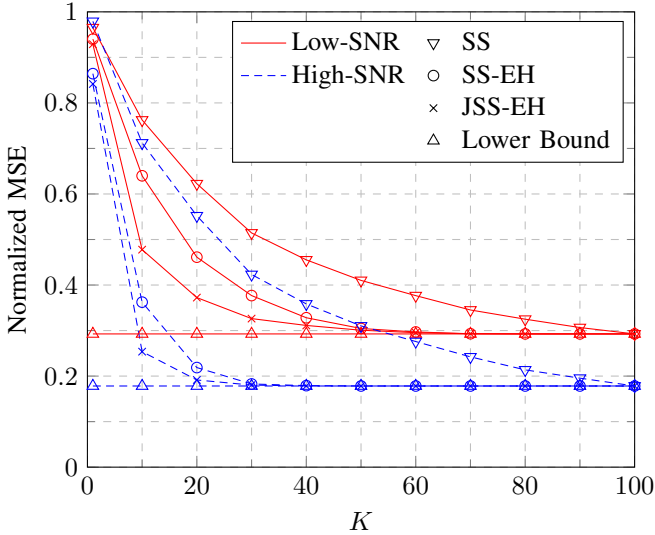}    
    \caption{Reconstruction distortion vs. number of active sensors, for high ($\sigma^2_w=0.01$) and low-SNR scenarios ($\sigma^2_w=0.5$) and lower bound ($M=100$, $T=20$, $\mu=0.25$).}
    \label{fig:MSEvsSS}
\end{figure}

Now, we focus our attention on the reconstruction distortion (MSE) for the proposed JSS-EH and SS-EH strategies (see Fig. \ref{fig:MSEvsSS}). For any strategy, a trivial Lower Bound (LB) of the optimal distortion can be found by letting $\mathcal{Z}_t=\mathcal{M}$ for all $t \in \mathcal{T}$ in problem \eqref{eq:OptRelaxed}. By doing so, we allow all sensors to be selected\footnote{Note this is not feasible since there are only $K\leq M$ orthogonal channels.} and, hence, all the observations can be used to reconstruct the source at the FC.

\begin{figure}[t]
    \centering
    \includegraphics[width=\columnwidth]{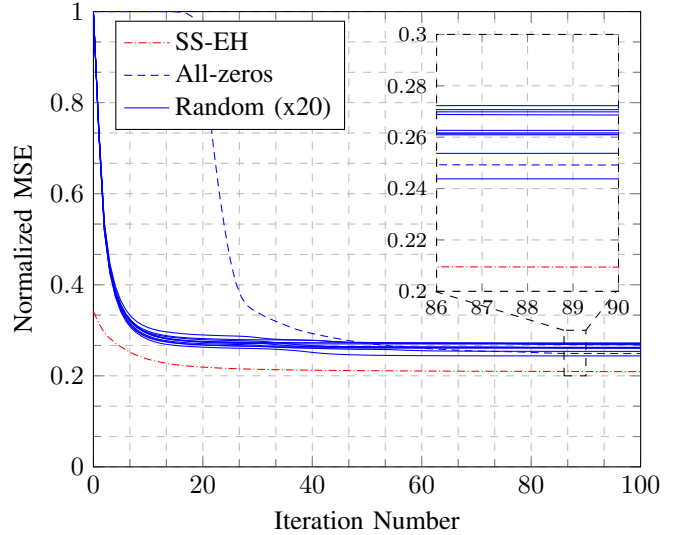}
    \caption{Reconstruction distortion vs. number of iterations for various initializations of the JSS-EH algorithm: SS-EH, all-zeros, and random for 20 different realizations ($M=50$, $T=20$, $K=10$, $\mu=1$, $\sigma_w^2=0.1$).}
    \label{fig:itJSS-EH}
\end{figure}

As expected, distortion monotonically decreases with $K$ in all cases. And, further, the resulting distortion is lower for the high-SNR scenario ($\sigma^2_w=0.01$). More importantly, the proposed JSS-EH and SS-EH strategies outperform the benchmark (SS), in particular for the high-SNR regime. Interestingly too, the gap between the JSS-EH curve and the lower bound is narrower than that of SS-EH for a low number of selected sensors, which turns out to be the region of interest. Also, for this scenario our proposed strategies attain the lower bound when the number of active sensors is set to $30\%$ and $50\%$, for the high- and low-SNR regimes, respectively. This implies that, yet suboptimal, the proposed JSS-EH and SS-EH strategies effectively attain the performance of the \emph{optimal} solution (which cannot be computed) when the number of active sensors is set to those values or higher. On the contrary, the benchmark SS strategy only attains the lower bound when all sensors are active.
Next, in Fig. \ref{fig:itJSS-EH}, we investigate the impact of the initialization on the performance (convergence rate, distortion after convergence) of the JSS-EH scheme. By far, the all-zeros initialization results into a slower convergence. Resorting to random initialization definitely helps speed up convergence. However, distortion can be further reduced by initializing the JSS-EH scheme with the solution to the SS-EH problem (including the resulting power allocation). This, in addition, results into faster convergence.

\subsection{Comparison of the Online and Offline Strategies}
\begin{figure}[t]
    \centering
    \includegraphics[width=\columnwidth]{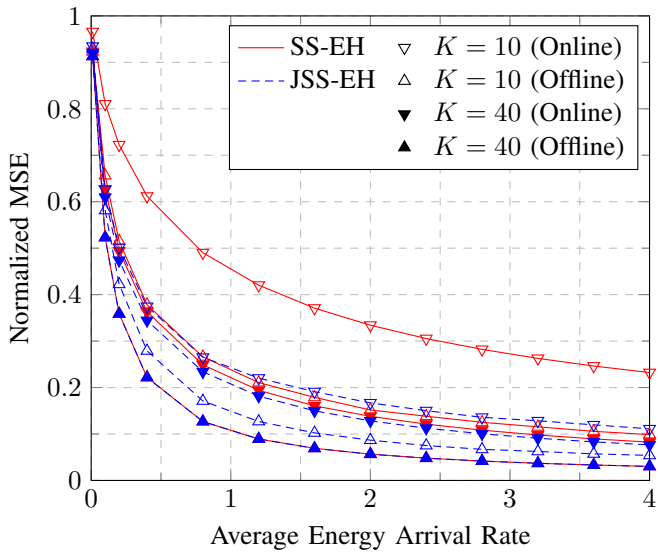}
    \caption{Distortion associated to the online and offline SS-EH and JSS-EH strategies, for scenarios with a low ($K=10$) and high ($K=40$) number of selected sensors ($M=50$, $T=20$, $\sigma_w^2=0.01$).}
    \label{fig:Online}
\end{figure}

Figure \ref{fig:Online} illustrates the performance of the offline and online strategies vs. the intensity rate of energy arrivals. Clearly, the distortion of the offline versions is lower and both the JSS-EH and SS-EH \emph{online} policies exhibit a similar behavior (yet distortion is lower for the former). To stress, distortion in all cases decreases for an increasing intensity rate of energy arrivals since, accordingly, the overall harvested energy increases too. Interestingly, in the SS-EH case, the gap between the online and offline curves is broader for a scenario with a low number of selected sensors ($K=10$ out of $M=50$, or $20\%$). However, this gap is particularly marginal for the JSS-EH scheme. For a conservative power allocation policy, if a substantial number of sensors with unspent energy are not scheduled in the final time slots, the remaining energy is wasted. And, clearly, this is more likely to happen for a lower number of selected sensors ($20\%$ vs. $80\%$).

To alleviate this, one can think of mechanisms to stimulate a more \emph{aggressive} (earlier) consumption of the harvested energy. For instance, rather than recomputing the solution for the remaining time slots, we can do so for a \emph{sliding window} of duration $T_w$, namely, for $t=t_o,\ldots, t_o+T_w$. The implicit assumption here is that no additional energy will be harvested in the few coming $T_w$ slots. By that time instant, the harvested energy should be consumed and, consequently, it favors an earlier consumption.

\begin{figure}[t]
    \centering
    \includegraphics[width=\columnwidth]{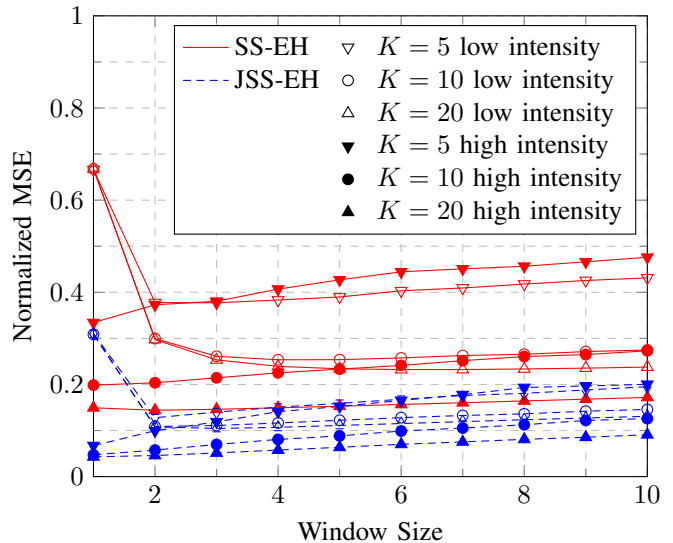}
    \caption{Reconstruction distortion for the online SS-EH and JSS-EH strategies, in scenarios with low ($\mu=0.1$, $E=25$) and high ($\mu=2.5$, $E=1$) intensity rates ($M=50$, $T=20$, $\sigma_w^2=0.01$).}
    \label{fig:OnlineSlidingWindow}
\end{figure}

In Figure \ref{fig:OnlineSlidingWindow}, we illustrate the impact of the window size ($T_w$) on the reconstruction distortion. Two different scenarios are considered: (i) low intensity rate, with high amounts of harvested energy in each arrival ($\mu=0.1$ and $E=25$); and (ii) high intensity rate, with low amounts of harvested energy ($\mu=2.5$ and $E=1$). In both scenarios, though, the average harvested energy is identical ($\mu \cdot E=2.5$). There exists a
trade-off in the duration of the sliding window $T_w$, as the curves for a low intensity rate of energy arrivals evidence. For very low $T_w$ values, energy is consumed shortly after being harvested (e.g., in the same time slot, for $T_w=1$). Consequently, transmission might need to be prematurely interrupted (i.e., $K$ might be larger than the number of sensors with available energy) which results into higher distortion. On the contrary, for high $T_w$ values (or when recomputing the solution for the whole remaining observation period), energy consumption is slower, which might result into some wasted energy in the final time slots (and, again, increased distortion). Therefore, there exists some intermediate (optimal) value yielding a minimum distortion (e.g., $T_w=2$ for $K=5$ in the SS-EH policy). Interestingly, the optimal duration of the sliding window becomes higher for an increasing number of selected sensors (namely, $T_w=4$ for $K=10$, $T_w=6$ for $K=20$ in the SS-EH policy). Intuitively, the risk of wasting energy when the percentage of scheduled sensors is higher turns out to be lower and, thus, sliding windows of a higher duration are advisable. For scenarios with high intensity rate ($\mu=2.5$, and $E=1$), curves are flatter. On the one hand, for low $T_w$ the risk of running out of energy before the next energy arrival is lower now and so is the distortion penalty (interestingly enough,
the optimal duration of the sliding window is one time slot, for $K=5$ and $K=10$). On the other hand, for high $T_w$ chances are lower that sensors remain unscheduled for a long time since energy arrives more frequently and the sensor selection and power allocation policies are more frequently recomputed too (lower distortion penalty again). Again, Figure \ref{fig:OnlineSlidingWindow} reveals a very similar behavior of the JSS-EH and SS-EH approaches for a varying window size (yet, unsurprisingly, distortion for the former is lower).  We also observe that the optimal window size tends to be smaller for the online JSS-EH policy.

\section{Conclusions}
\label{sec:Conclusions}

In this paper, we have proposed two suboptimal strategies to solve the non-convex problem of \emph{jointly} selecting a predefined number of energy-harvesting sensors and computing the optimal power allocation policy. The joint sensor selection and power allocation (JSS-EH) scheme is capable of finding a stationary solution (a proof is provided) on the basis of a majorization-minimization procedure. This allows us to identify a sequence of surrogate convex optimization problems that we iteratively solve. As an alternative, we propose a method to \emph{separately} identify a sensible sensor selection and power allocation policies (SS-EH scheme) which does takes into account the actual energy arrivals. The resulting power allocation strategy can be interpreted as a two-dimensional waterfilling solution.  We have also learned that the SS-EH solution turns out to be a suitable initialization to compute a \emph{decent} stationary solution to the JSS-EH problem in a relatively low number of iterations. The latter solution can be regarded as a \emph{refined} version with lower reconstruction distortion. Computer simulations revealed that the subsets of active sensors for the JSS-EH and SS-EH strategies are very similar. However, the corresponding power allocation policies differ. For the analyzed
scenario, the proposed strategies attain the lower bound when the number of active sensors is set to $30\%$ ($50\%$) in the high- (low-) SNR regime. We have also found that cropping the relaxed sensor selection vector of the JSS-EH scheme to the largest $K$ values without re-computing the power allocation policy has a negligible impact on distortion. Finally, we have proposed an \emph{online} version of the strategies. The associated distortion, however, is higher. This is in part motivated by the fact that it tends to generate \emph{conservative} power allocation patterns with slow energy consumption. Should a substantial fraction of those sensors not be scheduled by the end of the observation period, the harvested energy is wasted and, thus, distortion increases. By resorting to a sliding window, one can generate more \emph{aggressive} power allocation patterns (i.e., faster energy consumption). We have empirically shown that, for a given setting, an optimal duration of such sliding window exists (which might be in some cases, a single time slot).

\bibliographystyle{IEEEtran}
\bibliography{bib}
\end{document}